\documentclass[prl,twocolumn,showpacs,english,10pt,floatfix,groupedaddress,nofootinbib]{revtex4}
\usepackage{graphicx}
\usepackage{amssymb}
\usepackage{amsmath}
\usepackage{epsfig}

\makeatletter

\usepackage{pst-node}\usepackage{lscape}\usepackage{amsfonts}\usepackage{amsbsy}\usepackage{graphics}\usepackage{lscape}\usepackage{mathrsfs}\usepackage{dsfont}

\newcommand{\nn}{\nonumber}
\newcommand\as{\alpha_s}

\def \be  {\begin{equation}}
\def \ee  {\end{equation}}
\def \ba  {\begin{eqnarray}}
\def \ea  {\end{eqnarray}}
\def \baa {\begin{eqnarray*}}
\def \eaa {\end{eqnarray*}}

\def\slash#1{\setbox0=\hbox{$#1$}  
   \dimen0=\wd0     
   \setbox1=\hbox{/} \dimen1=\wd1  
   \ifdim\dimen0>\dimen1   
      \rlap{\hbox to \dimen0{\hfil/\hfil}} 
      #1     
   \else     
      \rlap{\hbox to \dimen1{\hfil$#1$\hfil}} 
      /      
   \fi}      %

\usepackage{babel}

\makeatother

\begin{document}

\title{Approximate next-to-next-to-leading order corrections to 
hadronic jet production}

\author{Daniel de Florian$^a$, Patriz Hinderer$^b$, Asmita Mukherjee$^c$,
Felix Ringer$^b$,  Werner Vogelsang$^b$}
\affiliation{$^a$Departamento de F\'{i}sica, FCEyN, Universidad de Buenos Aires,
(1428) Pabell\'{o}n 1, Ciudad Universitaria, Capital Federal, Argentina}
\affiliation{$^b$Institute for Theoretical Physics, 
T\"ubingen University, 72076 T\"ubingen, Germany}
\affiliation{$^c$Department of Physics, Indian Institute of Technology Bombay, 
Powai, Mumbai 400076, India}

\date{\today}

\begin{abstract}
We determine dominant next-to-next-to-leading order QCD corrections to single-inclusive jet
production at the LHC and Tevatron, using the established threshold resummation framework. 
In contrast to previous literature on this topic, our study incorporates all of the 
following features: (1) It properly accounts for the way a jet is defined in experiment
and treated in available full next-to-leading order calculations, (2) It includes the three 
leading classes of logarithmic terms in the perturbative expansion, and (3) It is adapted to the
full kinematics in jet transverse momentum and rapidity relevant for experiments. 
A recent full next-to-next-to-leading order calculation in the purely gluonic channel
allows us to assess the region where our approximate corrections provide an accurate
description. We expect our results to be important on the way to precision jet phenomenology 
at the LHC and as benchmark for further full next-to-next-to-leading order calculations. 
\end{abstract}

\maketitle

{\it{ Introduction.}}\,---\,The production of high-transverse-momentum hadron jets plays a fundamental 
role at the LHC~\cite{Meyer:2013kda} and at Tevatron~\cite{Mesropian:2011yn}. Jets are produced 
very copiously, making them precision probes of the physics of the Standard Model and beyond. 
Theoretical calculations whose precision matches that achievable in experiment are of critical 
importance. The efforts made in this context have spanned more than three decades now, culminating
so far with the recent calculation of the next-to-next-to-leading order (NNLO) perturbative corrections to 
jet production in the ``gluon-only'' channel~\cite{Ridder:2013mf,Currie:2013dwa}. 

As complete NNLO calculations of jet production are probably still a few years away, 
it is useful to determine approximate NNLO results, at least in certain kinematical regimes. 
This is possible thanks to the fact that the perturbative series for the partonic cross sections 
contains classes of logarithmic terms that often dominate. Resummation techniques in 
QCD~\cite{Kidonakis:1998bk} allow to determine the all-order structure of these logarithmic
terms, and one therefore also obtains the logarithms present at NNLO. Knowledge of
approximate NNLO expressions is very useful, since it potentially offers an avenue toward 
more precise phenomenology than available on the basis of the presently known full next-to-leading 
order (NLO) corrections. It also serves as benchmark for future full NNLO calculations. 

The logarithms just mentioned arise near a threshold from which the production of a jet becomes
possible in a partonic collision. They are hence known as ``threshold logarithms''. The threshold is
set by a vanishing invariant mass $\sqrt{s_4}$ of the partonic system that recoils against the observed 
jet. At the $k$th order of perturbation theory, one finds threshold corrections to the Born cross section
of the form $\as^k [\log^m(z)/z]_+$, with $0\leq m\leq 2k-1$, where $z=s_4/s$ with $\sqrt{s}$ 
the center-of-mass energy of the incoming partons. The systematic resummation of these logarithms
to all orders in the strong coupling $\as$ was derived for the case of jet production in~\cite{Kidonakis:1998bk}, 
where explicit next-to-leading logarithmic (NLL) results were given that in principle allow to resum the 
three ``towers'' of logarithms with $m=2k-1,2k-2,2k-3$. 

An important ``subtlety'' was pointed out in~\cite{Kidonakis:1998bk} concerning the 
threshold logarithms in jet production: the structure of the logarithmic corrections depends on 
whether or not the jet is assumed to be massless at partonic threshold, even at the leading-logarithmic
(LL) level. If the jet is taken to be massless at threshold, an approach for which we will use
the term ``scheme~(1)'' in the following, leading-logarithmic corrections arise
in the resummed perturbative function describing the jet. If, on the other hand, the jet is
permitted to have a non-vanishing invariant mass at threshold (``scheme~(2)''), the leading logarithms cancel,
leaving behind a non-leading logarithm whose coefficient depends on jet ``size'' parameter $R$ 
introduced by the jet algorithm. The difference between the two schemes may be understood from 
the fact that fewer final states contribute in scheme~(1) than in scheme~(2)~\cite{Kidonakis:1998bk}.

Approximate NNLO corrections for jet production have been derived 
in~\cite{Kidonakis:2000gi,Kumar:2013hia,Klasen:2013cba},
adopting scheme~(1). As one can see in the very recent study~\cite{Kumar:2013hia}, the NLO terms predicted 
for scheme~(1) fail to match a full NLO calculation~\cite{Cacciari:2011ma} even in a regime where
threshold logs are known to dominate. This becomes particularly evident from the fact that the threshold
terms for scheme~(1) do not carry any dependence on the jet parameter $R$, whereas the full NLO results
do. These features observed in~\cite{Kumar:2013hia} are in fact not surprising: explicit analytical NLO 
calculations~\cite{Jager:2004jh,Mukherjee:2012uz} have shown that jets produced close to partonic threshold 
do span a range of jet masses. Indeed, for any jet algorithm the jet produced in the perturbative calculation can
evidently contain two or more partons and hence have a non-vanishing invariant mass. This is even the case at 
exact threshold $z=0$, when for example only a single parton recoils against the entire jet. The maximally allowed
jet mass at threshold will depend on the parameter $R$ used in the jet algorithm. 

Thus, the assumption of massless jets at threshold that was made in previous 
studies~\cite{Kidonakis:2000gi,Kumar:2013hia,Klasen:2013cba} does not appear to be appropriate. 
Instead, the resummation ought to be carried out within scheme~(2). A resummed study
in this scheme was in fact performed in~\cite{deFlorian:2007fv}, where however only the
rapidity-integrated cross section was considered, for which the resummation simplifies
considerably. Integration over all rapidity is not quite adequate for comparisons with
experimental data. In the present paper we present new predictions for the NNLO threshold 
terms, using scheme~(2) and keeping full dependence on rapidity in the calculation. 
We will also go beyond the previous studies~\cite{Kidonakis:2000gi,Kumar:2013hia} by 
determining all three most leading logarithmic contributions $\propto (\log^3(z)/z)_+, (\log^2(z)/z)_+,
(\log(z)/z)_+$ at NNLO. The last of these is new; it may be obtained by matching the
resummation framework to a full NLO calculation. For the latter we choose that 
of~\cite{Jager:2004jh,Mukherjee:2012uz}, which provides analytical results for the partonic
cross sections. The calculation was performed assuming that the produced jet is rather narrow
(``narrow-jet approximation'' (NJA)). It has been shown that this approximation is
extremely accurate even at relatively large jet sizes of $R\gtrsim 0.7$. 

{\it Theoretical Framework.}\,---\,The factorized cross section for the single-inclusive
production of a jet with transverse momentum $p_T$ and pseudorapidity $\eta$ may be written as
\ba\label{crsec}
\frac{p_T^2 d^2\sigma}{dp_T ^2d\eta}&=&\sum_{ab}\int_0^{V(1-W)}dz
\int_{\frac{VW}{1-z}}^{1-\frac{1-V}{1-z}}dv\,x_a f_a(x_a,\mu_f)\nn\\
&&\hspace*{-0.49cm}
\times\, x_b f_b(x_b,\mu_f)\frac{d\hat\sigma_{{ab}}}{dvdz}(v,z,p_T,\mu_r,\mu_f,R),
\ea
where $V=1-x_T{\mathrm{e}}^{-\eta}/2$, $VW=x_T{\mathrm{e}}^{\eta}/2$, with
$x_T=2 p_T/\sqrt{S}$ and the hadronic center-of-mass energy $\sqrt{S}$. The sum runs
over all partonic collisions producing the jet; $d\hat\sigma_{{ab}}$ denote the corresponding
partonic hard-scattering cross sections and $f_a,f_b$ the parton distribution functions
at momentum fractions $x_a=VW/v(1-z)$, $x_b=(1-V)/(1-v)(1-z)$. The partonic cross sections
are computed in QCD perturbation theory. As indicated, besides depending on $p_T$ 
and the usual renormalization and factorization scales $\mu_r,\mu_f$, they are functions of 
the partonic kinematic variables, which we have chosen as
\be\label{crsec1}
v=\frac{u}{t+u}, \;\;\;\;\;\;\;\; z=\frac{s_4}{s},
\ee
where $s=x_ax_bS$ is the partonic center-of-mass energy squared, $t=(p_a-p_J)^2$, $u=(p_b-p_J)^2$ 
(with $p_{a,b}$ and $p_J$ the four-momenta of the initial partons and the jet, respectively),  
and $s_4$ is the invariant mass squared of the ``unobserved'' partonic system recoiling against the jet. 
We stress that the $d\hat\sigma_{{ab}}$ also depend on the algorithm adopted to define 
the jet, as indicated by the generic jet parameter $R$ in Eq.~(\ref{crsec}). We always assume
the jet to be defined by the anti-$k_t$ algorithm~\cite{Cacciari:2008gp}.

The perturbative series for each of the partonic scattering cross sections may be cast into the form
\be
\frac{sd\hat\sigma_{{ab}}}{dvdz}=\left(\frac{\alpha_s}{\pi}\right)^2\left[ 
\omega_{ab}^{(0)}+
\frac{\alpha_s}{\pi} \omega_{ab}^{(1)}+
\left(\frac{\alpha_s}{\pi}\right)^2 \omega_{ab}^{(2)}+{\cal O}(\alpha_s^3)\right],
\ee
where $\alpha_s\equiv\alpha_s(\mu)$ is the strong coupling constant, and 
where each of the $\omega_{ab}^{(k)}$ is a function of $v,z$ and, for $k>0$, of $R$ and
$p_T/\mu$ (we choose from now on $\mu_r=\mu_f\equiv \mu$). At lowest order we have 
\be
\omega_{ab}^{(0)}(v,z)\equiv\tilde\omega_{ab}^{(0)}(v) \delta(z),
\ee
since the recoiling system is a single massless parton. Hence $z=0$ sets a threshold for 
the process to take place, since the transverse momentum of the observed jet always needs 
to be balanced. At higher orders in perturbation theory, the hard scattering functions 
contain logarithmic distributions in $z$, with increasing powers of logarithms
as the perturbative order increases. More precisely, one has near the threshold at $z=0$:
\be
\alpha_s^k\omega_{ab}^{(k)}\,\sim\,\alpha_s^k\left(\frac{\log^m(z)}{z}\right)_+,\;\;\mathrm{with}\;\;
0\leq m\leq 2k-1.
\ee
Here
$\int_0^1 dzg(z)[f(z)]_+\equiv \int_0^1dz (g(z)-g(0))f(z)$. As one can see, two additional powers 
of the logarithm arise for every order of perturbation theory. Due to the integration against the
parton distribution functions, which are steeply falling functions of momentum fraction, the threshold
region $z\to 0$ typically makes significant contributions to the hadronic cross section. This is 
particularly the case when the kinematic boundary of the hadronic reaction is approached, that is, 
when $x_T\cosh\eta\to 1$. 

As is well known, the large logarithmic corrections arising in the threshold region are associated with
the emission of soft or collinear gluons. It is therefore possible to systematically determine the structure
of the corrections to all orders and to resum the ``towers'' of  logarithms with $m=2k-1, 2k-2,\ldots$.
This may be used to derive approximate beyond-NLO corrections for hadronic jet 
production, by expanding the resummed result appropriately to the desired 
order~\cite{Kidonakis:2000gi,deFlorian:2007fv,Kumar:2013hia}. To achieve the all-order 
resummation, one considers Mellin moments in $(1-z)$ of the partonic cross section:
\be
\Omega_{ab}(v,N)\equiv\int_0^1 dz (1-z)^{N-1}\frac{sd\hat\sigma_{{ab}}}{dvdz}.
\ee
In moment space, the resummed hard-scattering function $\Omega_{ab}^{\mathrm{res}}$ 
can at large $N$ be written as~\cite{Kidonakis:1998bk,Catani:2013vaa}
\ba
\Omega_{ab}^{\mathrm{res}}(v,N)&=&\sum_{c,d}
\Delta_a(N_a)\, \Delta_b(N_b)\,J^{\mathrm{(jet)}}_c(N,R)\,J^{\mathrm{(recoil)}}_d(N)\nn\\
&\times&\Delta_{ab\to cd}^{\mathrm{(int)}}(N,v) \,\Delta_c^{\mathrm{(ng)}}(N)  ,
\label{resumm}
\ea
where $N_a=vN$, $N_b=(1-v)N$ and the sum runs over the two final-state partons $c,d$
in an underlying $ab\to cd$ subprocess. Here it is assumed that parton $c$ produces the jet
(in a way that we shall clarify below), while the recoiling parton $d$ remains unobserved. 
Each of the terms is also a function of $\alpha_s(\mu)$ 
and $\log(\mu^2/s)$, which we have not written explicitly. 
Each of the functions $\Delta_a, \Delta_b,J_c^{\mathrm{(jet)}},J_d^{\mathrm{(recoil)}}$ 
is an exponential.  $\Delta_a, \Delta_b$ 
resum threshold logarithms arising from soft/collinear radiation off the incoming hard partons. 
Their expressions are very well known and may be found in the form we need them in, 
for example,~\cite{Catani:2013vaa}. Likewise, also the expression for gluon radiation 
off the ``unobserved'' recoiling parton $d$ is standard and may be found there. 
$\Delta_a, \Delta_b$ and $J_d^{\mathrm{(recoil)}}$ contain all the leading 
logarithmic pieces $\propto (\log^3(z)/z)_+, (\log^2(z)/z)_+$ in $\omega_{ab}^{(2)}$.

A crucial point of our study concerns the function $J_c^{\mathrm{(jet)}}$ used for 
the actual jet. As was shown in~\cite{Kidonakis:1998bk}, this function takes
different forms depending on whether one assumes the jet to become itself massless
at threshold or not. These two forms differ even at {\it leading} logarithmic level. 
For scheme~(2) introduced earlier, we have to 
next-to-leading logarithmic accuracy~\cite{Kidonakis:1998bk}:
\be
\log J_c^{\mathrm{(jet)}}= \int_s^{s/\bar{N}^2}\frac{dq^2}{q^2}
\,\as\left(q^2\right) \left(-\frac{C_c}{2\pi}\log\left(\frac{p_T^2 R^2}{s}\right)\right),
\ee
where $\bar{N}\equiv N {\mathrm{e}}^{\gamma_E}$ with the Euler constant 
$\gamma_E$, and where $C_c$ denotes the color charge of parton $c$, 
$C_q=C_F$ for a quark and $C_g=C_A$ for a gluon. As expected, $J_c^{\mathrm{(jet)}}$
is a function of $R$ in this scheme.

The function $\Delta_{ab\to cd}^{\mathrm{(int)}}(N,v)$ is obtained as a trace in color
space over hard, soft, and anomalous dimension matrices~\cite{Kidonakis:1998bk}.
All details have been given in~\cite{Kidonakis:2000gi} and need not be repeated
here. The function contributes at NLL level and is the only function in the resummed
expression that carries explicit dependence on $v$. 

Finally, $\Delta_c^{\mathrm{(ng)}}(N)$ in~(\ref{resumm}) contains the contributions from 
non-global logarithms. These were shown~\cite{Dasgupta:2001sh} to arise when an observable 
is sensitive to radiation in only a part of phase space, as is the case for a jet defined by some jet ``size'' 
parameter $R$. Their resummation is highly non-trivial. Non-global logarithms for
jet production first enter as a term $\propto [\log(z)/z]_+$ in $\omega_{ab}^{(2)}$.
As discussed in~\cite{Banfi:2010pa}, the non-global terms arise independently from the 
boundary of each individual (narrow) ``observed'' jet.
The appropriate second-order coefficient for our case of a single-inclusive jet cross 
section may therefore be directly obtained from~\cite{Dasgupta:2001sh,Banfi:2010pa}, adjusting
the argument of the logarithm properly. We note that these considerations-- and in fact the general 
structure of our resummed cross section-- apply to the anti-$k_t$ algorithm~\cite{Banfi:2010pa}.
We finally also mention that the non-global component 
makes a rather small contribution (a few per cent) to our numerical NNLO results presented below. 
All in all, after performing the Mellin-inverse to $z$-space, the two-loop expansion of 
the product $\Delta_{ab\to cd}^{\mathrm{(int)}}(N,v) \,
\Delta_c^{\mathrm{(ng)}}(N)$ in Eq.~(\ref{resumm}) takes the form 
\ba
&&\left(\frac{\alpha_s}{\pi}\right)^2\left[\tilde\omega_{ab}^{(0)}(v)\left(\delta(z)+\frac{1}{2}
\left(\frac{\alpha_s}{\pi}\right)^2{\cal C}^{\mathrm{(ng)}}_c\,\left(\frac{\log(z)}{z}\right)_+\right)\right.
\nn\\
&&\hspace*{16.3mm}+\frac{\alpha_s}{\pi}\left({\cal T}_{ab\to cd}(v)\delta(z)+
{\cal G}_{ab\to cd}^{(1)}(v)\left(\frac{1}{z}\right)_+
\right)\nn\\
&&\hspace*{16mm}+\left.\left(\frac{\alpha_s}{\pi}\right)^2{\cal G}^{(2)}_{ab\to cd}(v)
\left(\frac{\log(z)}{z}\right)_+\right],
\ea
with ${\cal C}^{\mathrm{(ng)}}_c=-C_AC_c\pi^2/3$ for the coefficient of the non-global term. 
The coefficients ${\cal G}_{ab\to cd}^{(1)}(v)$ are predicted by the resummation formalism.
The coefficients ${\cal T}_{ab\to cd}(v)$ may be derived by comparison to the explicit NLO results 
of~\cite{Mukherjee:2012uz} in the narrow-jet approximation. 
Along with the known resummation coefficients, knowledge 
of the ${\cal T}_{ab\to cd}(v)$ is sufficient for determining 
${\cal G}^{(2)}_{ab\to cd}(v)$~\cite{Kidonakis:2001nj,Almeida:2009jt}. In this way, combining with the 
contributions from $\Delta_a, \Delta_b,J_c^{\mathrm{(jet)}},J_d^{\mathrm{(recoil)}}$,
we obtain full control over the terms $\propto (\log^3(z)/z)_+, (\log^2(z)/z)_+,(\log(z)/z)_+$ 
in $\omega_{ab}^{(2)}$.
\begin{figure}[t]
\begin{center}
\vspace*{-2.9cm}
\hspace*{2mm}
\epsfig{figure=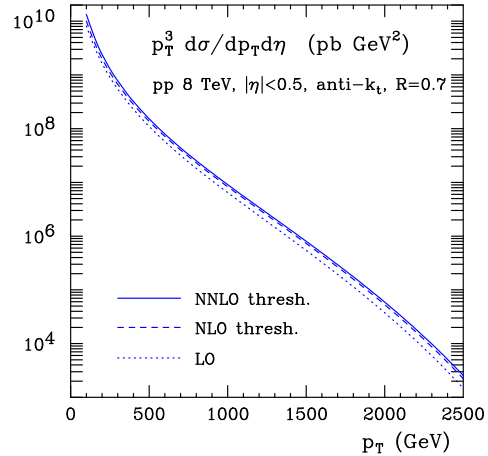,width=0.55\textwidth,angle=90}
\end{center}
\vspace*{-1.2cm}
\caption{{\it  Differential cross section for jet production in $pp$-collisions 
at the LHC at $\sqrt{S}=8$~TeV, using the anti-$k_t$ algorithm with $R=0.7$. 
\label{fig1} }}
\vspace*{0.cm}
\end{figure}

\begin{figure}[t]
\begin{center}
\vspace*{-2.1cm}
\hspace*{-8mm}
\epsfig{figure=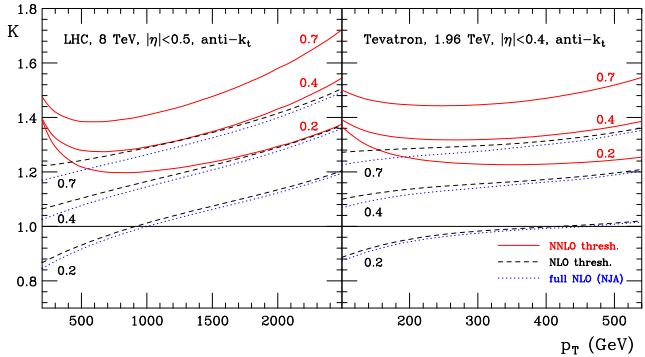,width=0.42\textwidth,angle=90}
\end{center}
\vspace*{-1.cm}
\caption{{\it  Left: $K$-factors for jet production in $pp$-collisions 
at the LHC at $\sqrt{S}=8$~TeV for $R=0.2,0.4,0.7$, using
the anti-$k_t$ algorithm. Right: Same for $p\bar{p}$ collisions at the
Tevatron at $\sqrt{S}=1.96$~TeV. \label{fig2} }}
\vspace*{0.cm}
\end{figure}

{\it Phenomenological results and discussion.}\,---\,Figure~\ref{fig1} shows results for the differential
single-inclusive jet cross section at the LHC, at lowest order as well as for the NLO and NNLO threshold terms. 
Here we use the CTEQ6.6~\cite{cteq66} parton distribution functions and scale $\mu=p_T$. The 
left part of  Figure~\ref{fig2} displays the corresponding ``$K$-factors'', defined as ratios of 
higher-order cross sections over the leading-order one, while the right part of the figure is 
for $p\bar{p}$ collisions at Tevatron at $\sqrt{S}=1.96$~TeV. Results are presented
for various jet parameters $R$. The dotted lines show
the NLO results of~\cite{Mukherjee:2012uz} which were obtained in the NJA for the anti-$k_t$ algorithm. 
We note that these agree with the NLO ones by the ``FastJet'' code~\cite{Cacciari:2011ma} (as shown 
in~\cite{Kumar:2013hia}) to better than $3\%$, even at $R=0.7$. The dashed lines present the results for
the NLO expansion of the threshold terms. It is evident that the latter provide a very faithful description of the 
full NLO results for much of the $p_T$ ranges relevant at LHC and Tevatron. This holds true for each value 
of $R$, thanks to the fact that the threshold logarithms carry $R$-dependence in our approach, in contrast to that
in~\cite{Kidonakis:2000gi,Kumar:2013hia}. Finally, the solid lines display the approximate NNLO
results. These show a striking further increase of the jet cross sections as compared to NLO, 
particularly so at high $p_T$ where the threshold terms are expected to dominate. 

\begin{figure}[t]
\begin{center}
\vspace*{-0.8cm}
\hspace*{-6mm}
\epsfig{figure=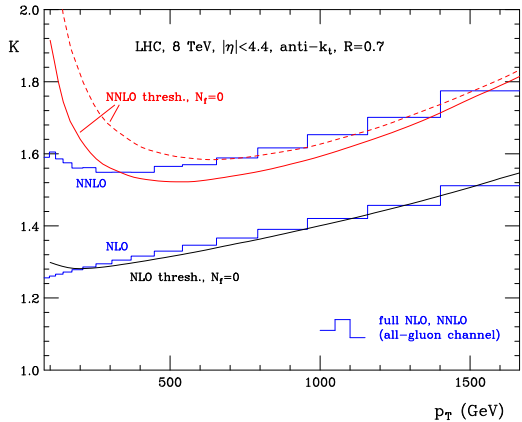,width=0.4\textwidth,angle=90}
\end{center}
\vspace*{-1.2cm}
\caption{{\it $K$-factors for jet production in $pp$-collisions 
at the LHC at $\sqrt{S}=8$~TeV in the ``gluon-only'' channel. The 
anti-$k_t$ algorithm with $R=0.7$ was used and the NNLO parton 
distributions of~\cite{Martin:2009iq}. The histograms show the results
of the recent full NNLO calculation~\cite{Currie:2013dwa} and its NLO 
counterpart, while the lines display the NLO and NNLO threshold terms. 
\label{fig3} }}
\vspace*{0.cm}
\end{figure}

Given the large size of the NNLO corrections observed in Fig.~\ref{fig2},
it is of course crucial to verify that the predicted enhancements are realistic. Fortunately,
recently a full NNLO calculation for jet production 
in the ``gluon-only'' channel was presented~\cite{Ridder:2013mf,Currie:2013dwa}, 
corresponding to $gg$ scattering and to setting the number of flavors $N_f=0$ in the 
partonic matrix elements. It is straightforward to compute our threshold 
terms in this limit. The comparison is shown in Fig.~\ref{fig3}. One can see
that the large enhancement at high $p_T$ predicted by the NNLO threshold terms 
is very nicely consistent with the full result. Judging from the comparison, the
NNLO threshold terms become accurate at about $p_T=400$~GeV for the chosen 
rapidity interval. Additional comparisons with the results of~\cite{Currie:2013dwa} 
show that this value is representative of rapidity intervals that contain the dominant region 
$\eta\approx 0$. 
One also finds that at very forward rapidities, $\eta\sim 4$,
our results indicate substantial NNLO $K$-factors of order 5 or so at $p_T\sim 40$~GeV.
This again appears to be consistent with the results shown in~\cite{Currie:2013dwa}.
In this regime, the coefficients of the threshold logarithms become large, due to ``small-$x$''
$t$-channel gluon exchange contributions. It will be important for future work to address this region in more
detail in order to derive reliable predictions for the forward jet cross section at the LHC.
Such contributions may also be responsible in part for the rise of the $K$-factor toward
lower $p_T$.
This rise is more pronounced for the NNLO threshold terms, implying
that subleading contributions become relevant here. Whether these are related to subleading
logarithmic terms, or to terms that vanish at partonic threshold $z=0$, will need to 
be studied in more detail. In order to shed light on terms of the latter type, the dashed line 
in Fig.~\ref{fig3} shows the NNLO threshold result found when using a different angular 
variable, $v'\equiv 1+t/s=z+v(1-z)$, in Eq.~(\ref{crsec}).  Clearly, $v'=v+{\cal O}(z)$.
The difference between the two NNLO threshold results
indicates a typical uncertainty of the prediction obtained from threshold resummation.

We thank A. Gehrmann-De Ridder, M. Klasen and S. Moch for useful communications.
We are grateful to M. Kumar for pointing out an inconsistency in our initial 
numerical results, and to J. Rojo for reporting detailed numerical studies obtained
with our code. AM thanks the Alexander von Humboldt Foundation, Germany, for 
support through a Fellowship for Experienced Researchers.



\begin{thebibliography}{99}
%

\bibitem{Meyer:2013kda} 
{\it See:} C.~Meyer [for the ATLAS and CMS Collaborations],
  arXiv:1310.2946 [hep-ex], {\it and references therein}.

\bibitem{Mesropian:2011yn} 
{\it See:} C.~Mesropian [CDF and D0 Collaborations],
  arXiv:1106.3119 [hep-ex], {\it and references therein}.

\bibitem{Ridder:2013mf} 
  A.~Gehrmann-De Ridder, T.~Gehrmann, E.~W.~N.~Glover and J.~Pires,
  Phys.\ Rev.\ Lett.\  {\bf 110}, 162003 (2013).
    
\bibitem{Currie:2013dwa} 
  J.~Currie, A.~Gehrmann-De Ridder, E.~W.~N.~Glover and J.~Pires,
  arXiv:1310.3993 [hep-ph]. {\it See the references for an extensive list
  of previous perturbative jet calculations.}
  
\bibitem{Kidonakis:1998bk} 
  N.~Kidonakis, G.~Oderda and G.~F.~Sterman,
  Nucl.\ Phys.\ B {\bf 525}, 299 (1998).
  
\bibitem{Kidonakis:2000gi} 
  N.~Kidonakis and J.~F.~Owens,
  Phys.\ Rev.\ D {\bf 63}, 054019 (2001).
  
\bibitem{Kumar:2013hia} 
  M.~C.~Kumar and S.~-O.~Moch,
  arXiv:1309.5311 [hep-ph].
  
\bibitem{Klasen:2013cba} 
  {\it See also:} M.~Klasen, G.~Kramer and M.~Michael,
  arXiv:1310.1724 [hep-ph].
  
\bibitem{Cacciari:2011ma} 
  M.~Cacciari, G.~P.~Salam and G.~Soyez,
  Eur.\ Phys.\ J.\ C {\bf 72}, 1896 (2012).

\bibitem{Jager:2004jh} 
  B.~J\"{a}ger, M.~Stratmann and W.~Vogelsang,
  Phys.\ Rev.\ D {\bf 70}, 034010 (2004).
  
\bibitem{Mukherjee:2012uz} 
  A.~Mukherjee and W.~Vogelsang,
  Phys.\ Rev.\ D {\bf 86}, 094009 (2012).
   
\bibitem{deFlorian:2007fv} 
  D.~de Florian and W.~Vogelsang,
  Phys.\ Rev.\ D {\bf 76}, 074031 (2007).
  
\bibitem{Cacciari:2008gp} 
  M.~Cacciari, G.~P.~Salam and G.~Soyez,
  JHEP {\bf 0804}, 063 (2008).
    
\bibitem{Catani:2013vaa} 
  S.~Catani, M.~Grazzini and A.~Torre,
  Nucl.\ Phys.\ B {\bf 874}, 720 (2013).
  
\bibitem{Dasgupta:2001sh} 
  M.~Dasgupta and G.~P.~Salam,
  Phys.\ Lett.\ B {\bf 512}, 323 (2001);
  A.~Banfi and M.~Dasgupta,
  JHEP {\bf 0401}, 027 (2004).

\bibitem{Banfi:2010pa}
A.~Banfi, M.~Dasgupta, K.~Khelifa-Kerfa and S.~Marzani,
  JHEP {\bf 1008}, 064 (2010).
  
\bibitem{Kidonakis:2001nj} 
  N.~Kidonakis, E.~Laenen, S.~Moch and R.~Vogt,
  Phys.\ Rev.\ D {\bf 64}, 114001 (2001).
  
\bibitem{Almeida:2009jt} 
  L.~G.~Almeida, G.~F.~Sterman and W.~Vogelsang,
  Phys.\ Rev.\ D {\bf 80}, 074016 (2009).
  
\bibitem{cteq66}
P.~M.~Nadolsky {\it et al.},
  Phys.\ Rev.\ D {\bf 78}, 013004 (2008).

\bibitem{Martin:2009iq} 
  A.~D.~Martin, W.~J.~Stirling, R.~S.~Thorne and G.~Watt,
  Eur.\ Phys.\ J.\ C {\bf 63}, 189 (2009).
    
\end{thebibliography}
\end{document}